\documentclass[11pt,twoside]{article}
\usepackage{newpasp}

\index{Churchill, C. W.}

\begin{document}

\title{Kinematics of log N({H\kern 0.1em {\small\bf
I}})$\simeq$17~cm$^{\rm\bf -2}$, Metal--Rich Gas Extended Around
Intermediate Redshift Galaxies}
\author{Chris Churchill}
\affil{The Pennsylvania State University} 


\begin{abstract}

The  kinematics  of  metal--enriched,  low--ionization  gas  at  large
galactocentric  distances are  compared to the host galaxy properties 
and are discussed in the context of global galaxy evolution.

\end{abstract}




\section{Galaxies Selected  by {Mg\kern 0.1em {\small\bf II}} Absorption}

Like  {\hbox{{\rm H}\kern 0.1em{\sc i}}}  studies,  the goal of quasar
absorption line studies is to develop an  understanding of the role of
gas in the formation and evolution  of galaxies.  Whereas an advantage
of {\hbox{{\rm H}\kern 0.1em{\sc i}}} studies is that both the spatial
and line--of--sight   velocity  distribution of   neutral gas   can be
studied in the  context of the  galaxy environment, quasar  absorption
lines   offer   a complementary  approach    using  an {\it  extremely
sensitive\/} ``pencil beam'' probe  through many galaxies over a  wide
range of cosmic epochs; independent of  redshift, absorption lines can
sample {\hbox{{\rm H}\kern  0.1em{\sc i}}} down to $\log N({\hbox{{\rm
H}\kern 0.1em{\sc i}}})  \simeq 12$~[{\hbox{cm$^{-2}$}}] (e.g.\ Tytler
1995).

Using the  resonant {{\rm Mg}\kern 0.1em{\sc ii}~$\lambda\lambda 2796,
2803$} doublet , {\hbox{{\rm  H}\kern 0.1em{\sc i}}}  can be probed in
galactic   environments  with $15   \leq  \log   N({\hbox{{\rm H}\kern
0.1em{\sc i}}}) \leq 21$~[{\hbox{cm$^{-2}$}}] (Churchill et~al.\ 1999;
Churchill  \&  Charlton   1999;   Churchill   et~al.\   2000),  making
{\hbox{{\rm Mg}\kern 0.1em{\sc ii}}}  an ideal  tracer for studies  of
global galaxy--gas evolution   (kinematic, chemical, and  ionization).
The   statistical  properties  of ``strongish''  {\hbox{{\rm  Mg}\kern
0.1em{\sc ii}}} systems  are thoroughly  documented to $z=2.2$  (e.g.\
Lanzetta, Turnshek,  \& Wolfe 1987; Steidel \&  Sargent  1992) and are
well established to be  ``associated'' with bright, normal galaxies of
almost all morphological types  to $z=1$ (e.g.\ Steidel, Dickinson, \&
Persson 1994;  Steidel et~al.\ 1997;   Steidel 1998).  Therefore, once
the requisite observational survey  work is undertaken, it is expected
that {\hbox{{\rm Mg}\kern 0.1em{\sc ii}}} absorption  can be used as a
tracer of extended gas around galaxies  to the highest redshifts where
stars first formed.

\section{Galaxies and Absorption Kinematics}

Because the {\hbox{{\rm H}\kern 0.1em{\sc i}}} column densities of the
gas probed in   {\hbox{{\rm  Mg}\kern 0.1em{\sc ii}}}  absorption  are
several orders of magnitude below those observed with 21--cm emission,
the regions and physical  conditions studied are quite different,  yet
complementary, to {\hbox{{\rm H}\kern   0.1em{\sc i}}} studies.    The
actual {\hbox{{\rm Mg}\kern 0.1em{\sc  ii}}} equivalent width measured
from a  parcel  of  gas   is  dependent upon   ionization   condition,
metallicity,  {\hbox{{\rm H}\kern  0.1em{\sc  i}}} column density, and
velocity dispersion.   For a metallicity of  0.1 solar, low ionization
gas  with  $\log     N({\hbox{{\rm H}\kern  0.1em{\sc    i}}})  \simeq
17$~[{\hbox{cm$^{-2}$}}] and  a line--of--sight velocity dispersion of
$\sim 20$~{\hbox{km~s$^{-1}$}} would  produce a rest--frame equivalent
width of $W_{r}(2796) \simeq 0.3$~{\AA}.   This  value is well  within
the    detection  capabilities    of  4--meter telescopes.    However,
dispersing the absorption at high  resolution, to reveal the component
to component     velocity  splittings,  requires  a    10--meter class
telescope.

In  Figure~1, WFPC2/{\it   HST\/}  images of 10   {\hbox{{\rm Mg}\kern
0.1em{\sc ii}}} absorption--selected galaxies (Steidel 1998) are shown
with corresponding absorption    line  profiles  of the    {\hbox{{\rm
Mg}\kern 0.1em{\sc ii}}} $\lambda 2796$  transition (Churchill \& Vogt
2000),  obtained with HIRES/Keck  I.   The galaxies  are presented  in
increasing   impact   parameter  order   from   left  to    right (the
sky--projected  separation between  the galaxy  center  and the quasar
line of sight).  The images are $2\arcsec \times 2\arcsec$; the galaxy
impact parameters are given in the upper portion of the panels and the
redshifts in the  lower portion.   The absorbing  gas is  shown in the
galaxy rest--frame velocity from  $-220$ to $220$~km~s$^{-1}$.   Ticks
above the profiles  give the numbers  and velocities of subcomponents,
based upon $\chi ^{2}$ Voigt profile fitting.

\begin{figure}[th]
\plotfiddle{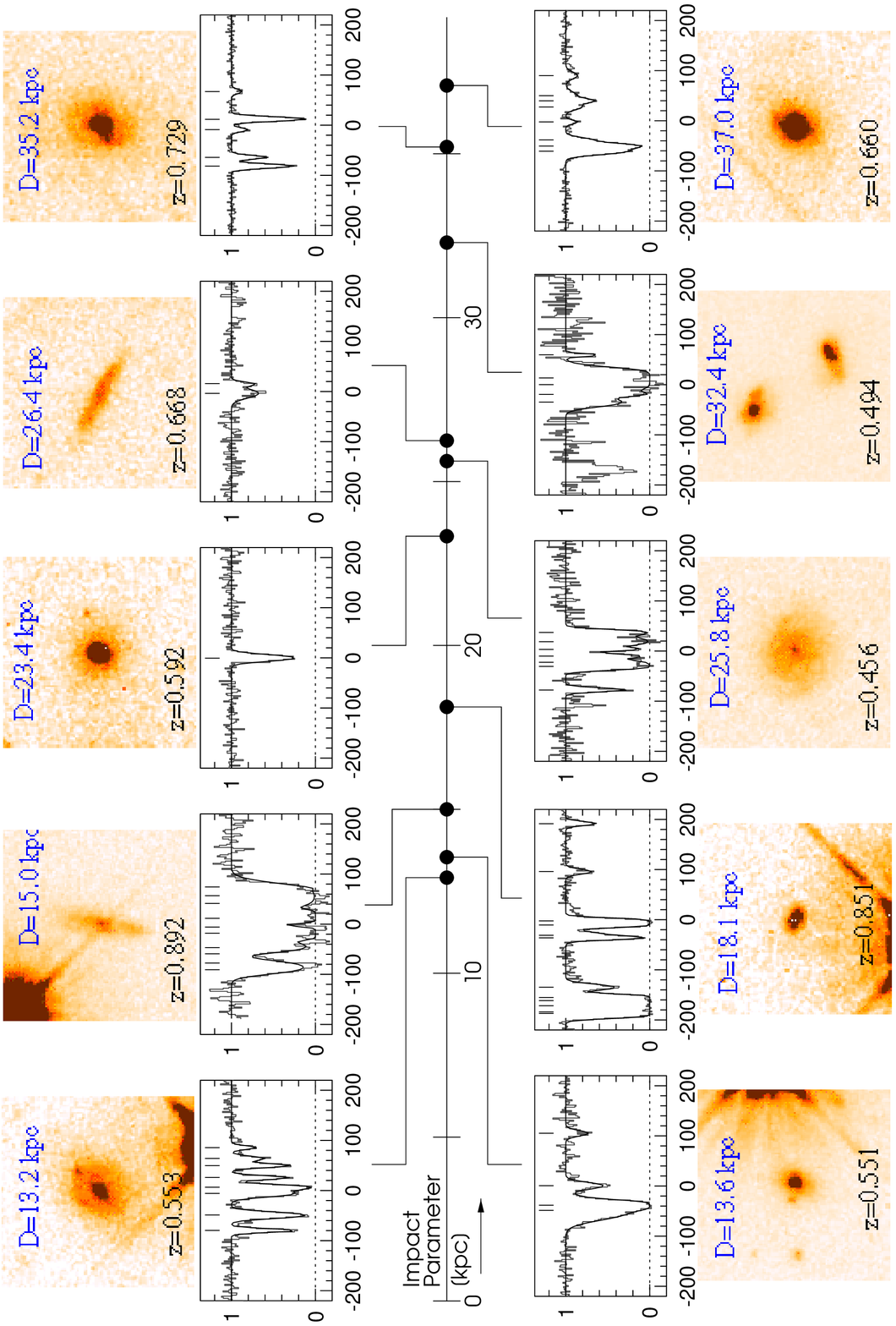}{4.0in}{-90.}{55.}{55.}{-205.}{290.}
\caption
{Galaxy morphology and  {\hbox{{\rm Mg}\kern 0.1em{\sc ii}}}  $\lambda
2796$ line--of--sight kinematics in  increasing impact parameter order
(see text for details).}
\end{figure}

These pencil beam probes reveal  a wide range of kinematic complexity,
with velocity  spreads of $\sim  150$~{\hbox{km~s$^{-1}$}}, at various
impact  parameters.  In small  samples,  including the one shown here,
there are   no statistically significant  ($> 3~\sigma$)  correlations
between   the   spatial   and     kinematic   distributions    of  the
metal--enriched, $N({\hbox{{\rm H}\kern 0.1em{\sc  i}}})$ gas  and the
galaxy  properties (Churchill, Steidel, \&   Vogt 1996).  However, the
stochastic   behavior in  the gas   properties  yield general  trends,
including decreasing equivalent  width (Steidel  1995), and increasing
ionization with decreasing   optical thickness with  impact  parameter
(Churchill et~al.\ 2000).

The absorption line  results are  consistent with {\hbox{{\rm  H}\kern
0.1em{\sc  i}}}  21--cm  kinematic  and spatial   distributions, which
extend to  several tens of kiloparsecs,  often in a flattened geometry
(e.g.\ Irwin  1995; also see Charlton \&   Churchill 1996, 1998).  The
main difference is that the {\hbox{{\rm  H}\kern 0.1em{\sc i}}} column
densities sampled  by the two techniques  differ  by as  much as three
orders of magnitude.  Within  a $\sim 40$~kpc region  around galaxies,
the covering  factor of  {\hbox{{\rm H}\kern  0.1em{\sc i}}}  gas with
$\log      N({\hbox{{\rm    H}\kern        0.1em{\sc      i}}})   \geq
19$~[{\hbox{cm$^{-2}$}}]  is    often patchy  and   highly  structured
spatially (see many contributions throughout this volume), whereas the
$\log N({\hbox{{\rm H}\kern 0.1em{\sc i}}}) = 17$~[{\hbox{cm$^{-2}$}}]
gas  has  a covering factor   very near unity (Steidel  et~al.\ 1994).
Regardless of the processes that gives rise  to the gas, it is evident
that the   filling  factor  (i.e.\  spatial  structure ``patchiness'')
decreases  as the {\hbox{{\rm  H}\kern  0.1em{\sc i}}} column  density
increases.

An additional important  point, that can be  gleaned from Figure~1, is
that what may appear as a ``normal'' galaxy in an optical image may be
far  from  normal  in its   spatial   and kinematic  distribution   of
metal--enriched gas.  An example is  the $z=0.851$ very blue,  compact
galaxy  at impact parameter 18.1~kpc  that  has velocity splittings as
large as 400~{\hbox{km~s$^{-1}$}} with complex variations in component
strengths.

\section{Absorption Lines and Global Galaxy Evolution}

Strong metal--line absorption  systems  are excellent  candidates  for
placing  powerful   constraints on global  galactic  evolution models.
Metals are  produced in stars, and  stars are produced in galaxies; an
$\alpha$--group element is best suited  since it  is  a tracer of  the
earliest stages   of stellar evolution   (i.e.\ Type  II  supernovae).
Also, the absorption lines should be accessible for study over a large
range  of  redshift (from $z\sim0$ to   $z\sim5$) so  that  gas can be
studied from its first enrichment by stars.   This requires a near--UV
transition so that confusion with {\hbox{{\rm Ly}\kern 0.1em$\alpha$}}
forest  lines   can   be   avoided.  The   {{\rm  Mg}\kern   0.1em{\sc
ii}~$\lambda\lambda 2796, 2803$} doublet meets all these criteria.

\begin{figure}[hbt]
\plotfiddle{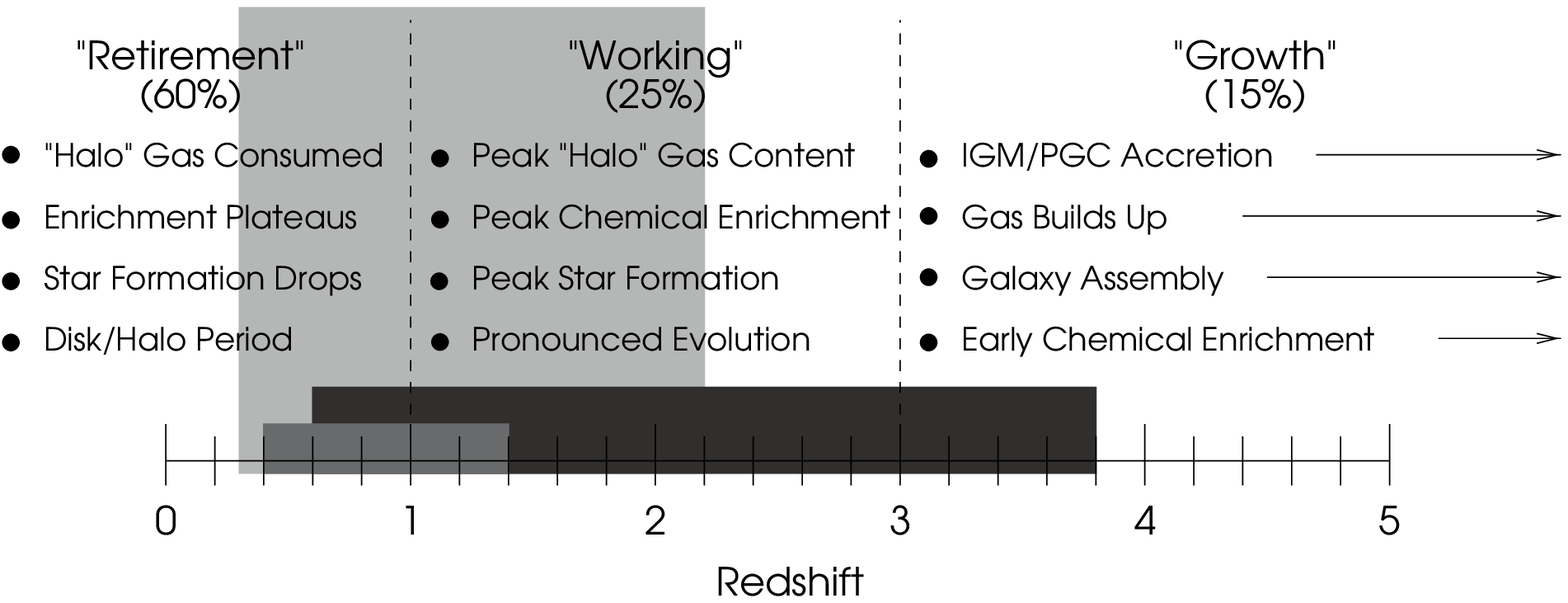}{2.0in}{0.}{70.}{70.}{-200.}{0.}
\caption
{Schematic  of the  Pei, Fall, \&   Hauser  (1999) scenario of  global
galaxy evolution and how it compares  to observed {\hbox{{\rm Mg}\kern
0.1em{\sc ii}}} redshift and sensitivity coverage (greys) and proposed
future coverage (black).}
\end{figure}

Figure~2 is  a  schematic of global  galaxy  evolution based  upon the
models of Pei, Fall, \& Hauser (1999).  Three  basic epochs are shown:
``Growth''  ($z > 3$), ``Working''   ($1<z\leq 3$), and ``Retirement''
($z\leq  1$),  with the     respective percent   of  cosmic time    in
parenthesis.    As shown with     grey shading, {\hbox{{\rm   Mg}\kern
0.1em{\sc ii}}} surveys  have been conducted  with detection limits of
$0.3$~{\AA} from $0.3 \leq z \leq 2.2$ (Lanzetta et~al.\ 1987; Steidel
\&  Sargent 1992)  and  of  $0.02$~{\AA}  (with high  resolution) from
$0.4\leq z \leq 1.4$ (Churchill et~al.\ 1999; Churchill \& Vogt 2000).

There is  strong  differential evolution  in the  {\hbox{{\rm Mg}\kern
0.1em{\sc   ii}}}  equivalent  width  distribution    from $z\sim2$ to
$z\sim0.5$ in that the strongest systems  evolve away from ``Working''
epoch to the ``Retirment''   epoch.  Observationally, this is   due to
differential evolution in the complexity  of the {\hbox{{\rm  Mg}\kern
0.1em{\sc ii}}} kinematics with equivalent width (Churchill
\& Vogt 2000).  Physically, it is likely due to strong metallicity and
stellar evolution until $z=1$.  An additional  clue to the {\hbox{{\rm
Mg}\kern 0.1em{\sc ii}}} kinematics  evolution is a strong correlation
of a separate ``{\hbox{{\rm  C}\kern 0.1em{\sc iv}}} phase'' with  the
{\hbox{{\rm Mg}\kern   0.1em{\sc  ii}}} kinematics  (Churchill et~al.\
2000).  This suggests a direct connection between mechanisms that give
rise to  extended, multiphase  halos,   e.g.\ star formation   (Dahlem
1998), and {\hbox{{\rm Mg}\kern 0.1em{\sc ii}}} kinematics.

It is  expected that the properties  of {\hbox{{\rm Mg}\kern 0.1em{\sc
ii}}} systems will  evolve even  more  dramatically from  $z=4$, which
includes the   ``Growth'' phase (observations  require high resolution
infrared spectroscopy on large aperture telescopes).



\acknowledgements
Thanks  are due to J. Charlton,  B. Jannuzi, R.  Mellon, J. Rigby, and
C. Steidel for their contributions to the work presented here.

\end{document}